\documentclass[pra,twocolumn,superscriptaddress,showpacs,amssymb]{revtex4}
\usepackage[dvipdfm]{graphicx}

\begin{document}
\title{Vortex Ring Dynamics in Trapped Bose-Einstein Condensates}

\author{Matthew D. \surname{Reichl}}
	\affiliation{Laboratory of Atomic and Solid State Physics, Cornell University, Ithaca, New York 14853, USA}
	
\author{Erich J. \surname{Mueller}}
	\affiliation{Laboratory of Atomic and Solid State Physics, Cornell University, Ithaca, New York 14853, USA}

\date{\today}


\begin{abstract}        % give a summary of your paper
 We use the time-dependent Gross-Pitaevskii equation to study the motion of a vortex ring produced by phase imprinting on an elongated cloud of cold atoms. Our approach models the experiments of Yefsah et. al. [Nature \textbf{499}, 426] on $^6$Li in the BEC regime where the fermions are tightly bound into bosonic dimers. We find ring oscillation periods which are much larger than the period of the axial harmonic trap. Our results lend further strength to Bulgac et. al.'s arguments [arXiv: 1306.4266] that the ``heavy solitons" seen in those experiments are actually vortex rings. We numerically calculate the periods of oscillation for the vortex rings as a function of interaction strength, trap aspect ratio, and minimum vortex ring radius. In the presence of axial anisotropies the rings undergo complicated internal dynamics where they break into sets of vortex lines, then later combine into rings. These structures oscillate with a similar frequency to simple axially symmetric rings. 
%                         please supply keywords within your abstract
\end{abstract}
\maketitle

\section{Introduction}

Yefsah et. al. \cite{yefsah2013} recently observed anomalously slow oscillations of a nominal soliton in a harmonically trapped fermonic superfluid. By illuminating half the cloud with light, they generated a phase profile with a large jump. This phase jump evolved into a localized density depletion that oscillated with a period many times larger than the period associated with the harmonic trap. This slow motion is remarkable, as it exceeds the best theoretical calculations of the oscillation frequency of a soliton \cite{liao2011, scott2011} by an order of magnitude. Recently, Bulgac et. al. \cite{bulgac2013} hypothesized that the experimental protocol produces a vortex ring instead of a soliton. Through integrating time dependent equations based upon a superfluid density functional theory, Bulgac et. al. showed that near unitarity the experimental observations are consistent with this vortex ring hypothesis. Here we extend this analysis to the BEC regime, where the fermions are tightly bound into dimers. 

To study this problem, we numerically evolve the time-dependent Gross-Pitaevskii (GP) equation to simulate the dynamics of vortex rings. We model the phase imprinting process and measure the period of oscillations of the vortex ring as a function of interaction strength, trap aspect ratio, and initial ring radius. We find that the period of oscillation for the vortex ring is quantitatively consistent with the experimental observations in the BEC regime. 

In our simulations, the phase imprinting produces a soliton \cite{frantzeskakis2010}  which decays into a vortex ring through a ``snake instability". This instability has been explored in the past \cite{kuznetsov1995, muryshev1999, feder2000, frantzeskakis2010, cetoli2013}, as has the structure and motion of individual vortex rings \cite{shariff1992, batchelor2000, rayfield1964, amit1966, roberts1971, jones1982, donnelly1991, koplik1996, jackson1999, guilleumas2002, hsueh2007, abad2008,  caplan2012}. Despite these precedents, our exploration of this full process, in this particular geometry, is novel. An excellent review of the theory of solitons and vortices in BEC's can be found in Ref.~\cite{carretero2008}.

Experimentally, vortex rings have been observed in the decay of dark solitons \cite{anderson2001}, in complex vortex ring/soliton oscillations \cite{ginsberg2005, shomroni2009}, and in binary condensates \cite{mertes2007} . If Bulgac's \cite{bulgac2013} hypothesis is correct (as it as it appears to be) the experiment by Yefsah et. al. \cite{yefsah2013} can be added to this list and is the first experimental realization of a vortex ring in a fermionic superfluid.

\section{Physics of Vortex Rings} \label{sectheory}

The flow in a Bose condensate is irrotational ($\nabla \times \mathbf{v}=0$, where $\mathbf{v}= \frac{\hbar}{m} \nabla \phi$ is the local velocity and $\phi$ is the phase of the order parameter) except at line singularities. The superfluid phase winds by $2 \pi n$, for integer $n$, when one moves around one of these vortex lines. Here we study configurations where these vortex lines form loops. In particular, consider a cigar shaped cloud, elongated along the $\hat{z}$ axis, with a vortex ring in a perpendicular plane, concentric with the cloud. In Sec.~\ref{secnum} we numerically solved the time-dependent GP equation to analyze such a ring, but its basic properties can be understood from a semiclassical argument given by Jackson et. al. \cite{jackson1999} for a vortex ring in a spherically symmetric condensate. They find that each element of the vortex ring moves with a velocity $\mathbf{v}$:

\begin{equation}
\mathbf{v}=v_{in}(R)\hat{\mathbf{z}}+ \omega_p \hat{\mathbf{\kappa}} \times \mathbf{r}
\label{scmod}
\end{equation}
where $\hat{\kappa}$ defines the direction of the circulation at the element. 

The first term in Eq.~\ref{scmod} describes the induced velocity $v_{in}(R)$ resulting from the sum of the velocity contributions from each element on the ring. For a ring in a uniform condensate, this induced velocity is a function of the ring radius $R$ and is given by $v_{in}(R)= \frac{\hbar }{2m R} [\log(8R/\xi)-0.615]$ \cite{roberts1971}, where $\xi$ is the coherence length. Thus the ring has an inherent tendency to propagate along the z-axis. 

A cartoon of this physics comes from the two-dimensional analog of a vortex ring: a vortex dipole consisting of two point vortices--one with circulation $+\kappa$, the other with $-\kappa$. If these are separated by a distance $2R$, they move with a velocity $v= \frac{\kappa}{4\pi R}$ \cite{pelinovsky2011}. 

The second term in Eq.~\ref{scmod} describes the Magnus force on the ring caused by the harmonic trap. In the case of a straight vortex line, this forces leads to precession with frequency $\omega_p$. Note that $w_p $ is not equal to any trap frequency; for instance, in the Thomas-Fermi limit, a single vortex in a two-dimensional condensate will precess with a frequency given by~\cite{fetter2001}

 \begin{equation}
 \omega_p= \frac{3 \hbar \omega^2}{4 \mu} \log (\frac{R_{\perp}}{\xi})
 \label{precfreq}
 \end{equation}
 where $\omega$ is the trap frequency, $\mu= \hbar^2/2m \xi^2$ and $R_{\perp}^2= 2\mu/ m\omega^2$.  

A small ring ($R<<R_\perp$) beginning at $z=0$ will have a large velocity component in the positive $z$ direction. As it moves in the $z$ direction the Magnus term, $\hat{\kappa} \times \mathbf{r}$, causes the ring to grow. Once the ring radius is sufficiently large the Magnus force pushes the ring in the negative $z$ direction. In this manner the ring oscillates. The two dimensional analog of this motion was observed in experiments by Neely et. al. \cite{neely2010}.

While this model is too simple to produce a quantitative prediction for the period of these vortex ring oscillations, it captures the qualitative behavior of the vortex ring seen in the numerical simulations discussed in Sec.~\ref{secosc}. Moreover, it predicts that the period should increase roughly as $T_{ring} \sim \frac{1}{\xi^2\log(1/{\xi})} \sim gn/\log{gn} \sim g^{2/5}/\log{g}$ where $g$ is the interaction strength and $n$ is the density. This scaling is seen in our simulations (see Fig.~\ref{pervg}). We note that the period of dark soliton oscillations, $T_{sol}= \sqrt{2} T_z$ \cite{busch2000, konotop2004, becker2008} where $T_z$ is the trap period in the $z$ direction, is independent of $g$. For sufficiently large $g$ we expect slower oscillations for the vortex ring than the dark soliton.  

We also note that the vortex ring dynamics are highly nonlinear, and that the ring's oscillation period is strongly dependent on the minimum ring radius. As in Bulgac et. al.'s work \cite{bulgac2013}, we find that smaller radii lead to shorter periods (see Fig.~\ref{pervrmin}). We also find that the oscillations are non-sinusoidal, with a slight asymmetry between the motion to the left and to the right.

\section{Numerical Results}\label{secnum}

\subsection{Simulation Details} \label{secsim}
In this section we present results from numerical simulations of the time-dependent GP equation: 
\begin{equation}
i \hbar \partial_t \psi= -\frac{\hbar^2}{2m} \nabla^2 \psi + V_t (r,z) \psi +\frac{4 \pi \hbar^2 a N}{m} |\psi|^2 \psi + V_i (t, r, z) \psi
\end{equation}
where $N$ is the total number of particles, $a$ is the scattering length and $\psi$ is normalized such that 
\begin{equation}
\int |\psi(\vec{r})|^2 d^3 r= 1
\end{equation}
$V_t(r,z)=  \frac{1}{2}(\omega_r^2 r^2+ \omega_z^2 z^2)$ is a harmonic trapping potential and $V_i(t,r,z)$ is a time dependent phase-imprinting potential which we'll describe below. 

After rescaling the variables $t \rightarrow \omega_z t $, $\vec{r} \rightarrow \frac{1}{a_z}  \vec{r}$ where $a_z= \sqrt{\frac{\hbar}{m \omega_z}}$, and rescaling   
$\psi \rightarrow \frac{1}{a_z^{3/2} } \psi$, we can rewrite the GP equation in the dimensionless form:
\begin{equation} 
i \partial_t \psi = -\frac{1}{2} \nabla^2 \psi + g |\psi|^2 \psi + \frac{1}{2} (\lambda^2 r^2+z^2)\psi + V_i (t, r, z) \psi
\label{dgpe}
\end{equation}
where $\lambda= \frac{\omega_r}{\omega_z}$ is the trap aspect ratio and $g= \frac{4 \pi a N}{a_z}$ parametrizes the interaction strength. As discussed in Sec.~\ref{secosc}, an experimentally relevant set of parameters are $\lambda= 6$ and $g= 3 \times 10^4$.

In Secs.~\ref{secex} and \ref{secosc} we assume axial symmetry while in Sec.~\ref{secasym} we carry out full 3D simulations, including slight trap asymmetries. We numerically solve Eq.~\ref{dgpe} using a split-step method. We use a square grid, choosing our grid spacing sufficiently small that the dynamics are independent of the grid. We find for our parameter range that it suffices to take $\delta r= \delta z = 0.1$. Smaller grids are necessary for larger interactions. Similarly we find a time step $\delta t= 10^{-3}$ suffices for preventing large phase jumps between time steps, ensuring numerical stability. We set  $V_i (t, r, z) = \frac{\pi}{\delta t} \Theta (t) \Theta(\delta t- t) f(z)$ so that a sharp $\phi$ phase jump is imprinted about the line $z=0$ after the first time step. Each simulation begins after first relaxing the system into the ground state of the trapping potential using imaginary time propagation. 

The resulting dynamics after phase-imprinting can depend sensitively on the precise shape of $f(z)$. However, away from the quasi-one dimensional regime ($\lambda << n g$),  we find from our simulations that $f(z)$ generically creates a soliton that quickly decays into one (as in Fig.~\ref{solun}) or more vortex rings via a snake instability. For simplicity, and in keeping with the experimental observations in Ref.~\cite{yefsah2013} where there is only one discernible density depletion, we choose $f(z)= \frac{1}{2}(1+\tanh(z/\delta z))$  where $\delta z$ is our numerical grid spacing. This protocol consistently results in only one long lasting vortex ring. 

It is difficult to control the minimum radius $R_{min}$ of the vortex ring using this phase printing technique. Therefore, to study the behavior of the vortex ring as function of $R_{min}$, we do not use phase imprinting but instead relax the gas to a state with the following ansatz for its phase:
\begin{equation}
\label{phaseeq}
\frac{\psi(r,z)}{|\psi(r,z)|}= \frac{(r-R_{min})+i z}{\sqrt{(r-R_{min})^2+z^2}}
\label{ansatz}
\end{equation}
This ansatz closely approximates the phase of the vortex rings created after phase imprinting and allows us to easily investigate the ring behavior as a function of $R_{min}$. 

\subsection{Example of Snake Instability and Vortex Ring Dynamics} \label{secex}
\begin{figure} 

\hbox{\hspace{-5.0em}
\includegraphics[width=0.6\textwidth]{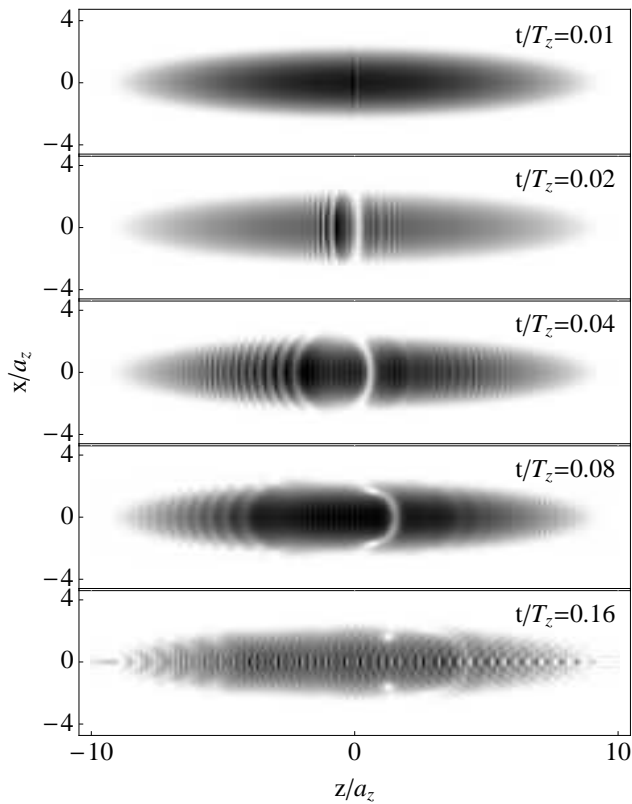}}
\caption{Condensate dynamics following phase imprinting with $g= 4000$, $\lambda=4$. Each graph shows the density $|\psi(x, y=0, z)|^2$, where darker colors represent higher density.  The dark soliton is unstable and forms a vortex ring (seen as two zero density cores) at time $t/T_z \approx 0.16$} 
\label{solun}
\end{figure}

Fig.~\ref{solun} shows an example of the dynamics of the condensate following phase imprinting. A soliton, seen as a density dip extending axially through the condensate, travels in the positive $z$ direction and almost immediately begins bowing outward near the center of the gas (see $t/T_z=0.04$ in Fig.~\ref{solun}). By time $t/T_z=0.16$ the soliton has decayed via this snake instability leaving a vortex ring which is seen as two zero density cores in the $y=0$ slice shown in Fig.~\ref{solun}.

\begin{figure}
\hbox{\hspace{-5.0em}
\includegraphics[width=0.6\textwidth]{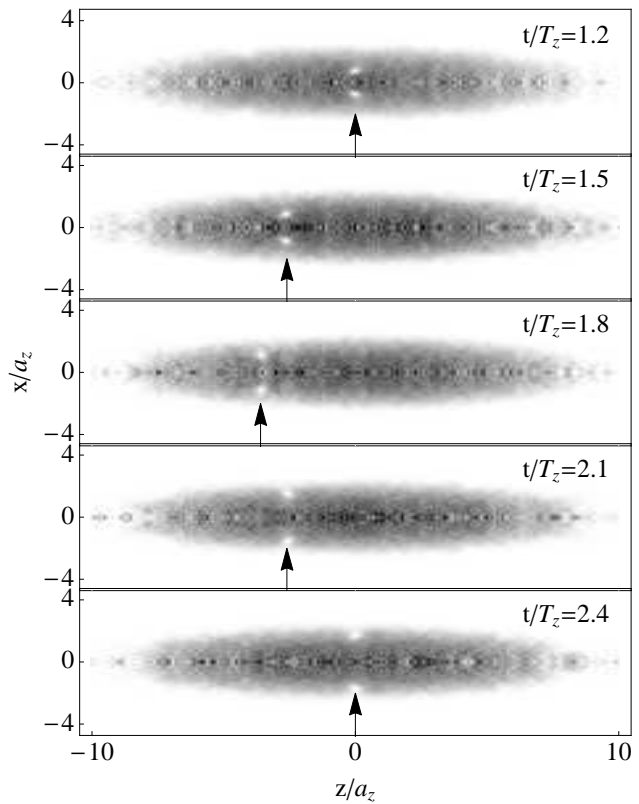}}
\caption{Vortex ring oscillation with  $g= 4000$, $\lambda=4$. Each graph shows the density $|\psi(x, y=0, z)|^2$, where darker colors represent higher density. An arrow is shown pointing to the vortex ring. } 
 \label{ringo}
\end{figure}

Fig.~\ref{ringo} shows an example of the vortex ring oscillations that follow the decay of a soliton. At $t/T_z \approx 1.2$ the vortex ring is positioned at $z=0$ and is traveling in the negative $z$ direction. The ring continues to travel in this direction until $t/T_z \approx 1.8$. After this time the ring radius expands to the edge of condensate while the ring begins traveling back in the positive $z$ direction. The ring completes half of an oscillation and returns to $z=0$ at $t/T_z \approx 2.4$. In the following subsection we calculate the frequency of vortex ring oscillations as a function of $g$, $\lambda$, and $R_{min}$.

\subsection{Period of Vortex Ring Oscillations} \label{secosc}
\begin{figure}
\centering
\includegraphics[width=0.45\textwidth]{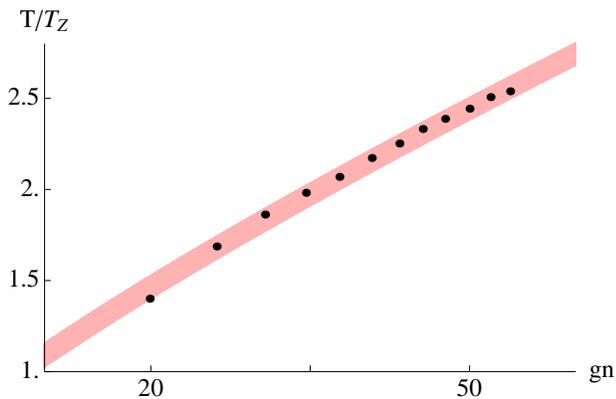}
\caption{Vortex ring oscillation period $T$ (normalized by the trap period $T_z$)  versus interaction strength times the density $gn$ with a trap aspect ratio of $\lambda = 4$ after phase imprinting. As predicted in Sec.~\ref{sectheory}, $T\sim gn/ \log{gn}$; the thick red curve shows a fit of the data to this scaling. } 
\label{pervg}
\end{figure}

Fig.~\ref{pervg} shows a plot of the vortex ring oscillation period as a function of interaction strength $g$ with a trap aspect ratio of $\lambda=4$ Each point is computed by first preparing the vortex ring with the phase imprinting method discussed in Sec.~\ref{secsim}, and then calculating the number of time steps for a vortex core starting at $z=0$ to complete an oscillation and return to $z=0$. As predicted above, the oscillation period increases as $gn/\log gn$ for large $gn$. Moreover, for $g \gtrsim 500$, the vortex ring oscillates at a period larger than the period for a dark soliton in a BEC $T= \sqrt{2} T_z$ \cite{busch2000, konotop2004, becker2008}. 

\begin{figure}
\centering
\includegraphics[width=0.45\textwidth]{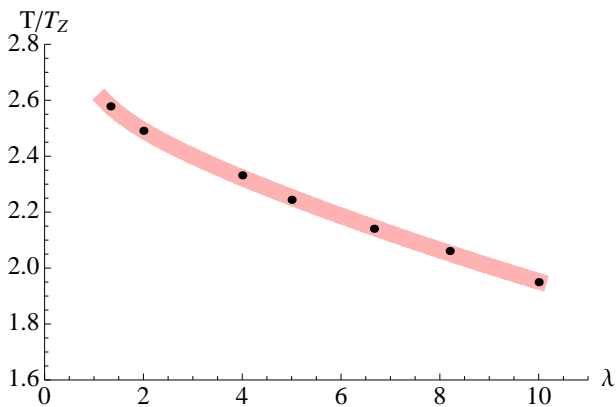}
\caption{Vortex ring oscillation period $T$ (normalized by the trap period $T_z$)  versus trap aspect ratio $\lambda= \frac{\omega_r}{\omega_z}$ after phase imprinting with $g=4000$. The thick red curve is an interpolation to guide the eye.} 
\label{pervtrap}
\end{figure}

In Fig.~\ref{pervtrap} we plot the oscillation period of the ring as function of trap aspect ratio $\lambda$, at constant interaction strength $g=4000$. The period decreases at larger aspect ratios which is consistent with observations in Ref.~\cite{yefsah2013}. The explanation for this trend is that our phase imprinting method yields vortex rings with smaller minimum radii at larger $\lambda$. As discussed in Sec.~\ref{sectheory}, the rings with smaller $R_{min}$ have smaller periods. 

\begin{figure}
\centering
\includegraphics[width=0.45\textwidth]{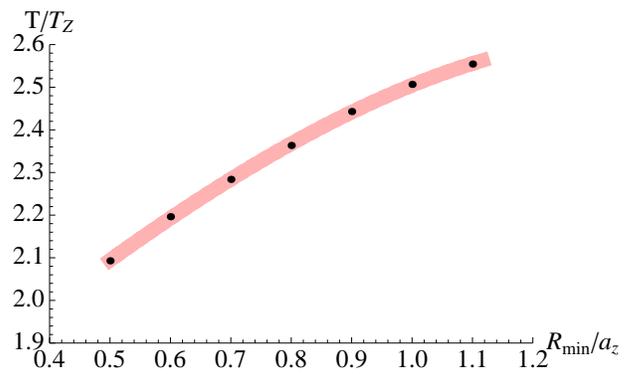}
\caption{Vortex ring oscillation period $T$ (normalized by the trap period $T_z$)  versus the minimum vortex ring radius $R_{min}$ with $g=4000$ and $\lambda= 4$. The thick red curve is an interpolation to guide the eye.} 
\label{pervrmin}
\end{figure}

To explore the radius dependence of the ring dynamics, we find the ring oscillation period as function of $R_{min}$ (see Eq.~\ref{phaseeq}) with $g=4000$ and $\lambda=4$ using the relaxation procedure discussed in the last paragraph of Sec.~\ref{secsim}. The results are shown in Fig.~\ref{pervrmin} which clearly demonstrates that rings with smaller $R_{min}$ have smaller periods; this is consistent with a similar finding reported in Ref.~\cite{bulgac2013}.

Finally, we compare our simulations to the experiment in Ref.~\cite{yefsah2013}. Typical experimental parameters in the BEC regime are: $T_r \approx 14$ms, $T_z \approx 87$ms, total number of bosonic Feshbach molecules $N\approx 1.1 \times 10^5$, Thomas-Fermi radius $R_{TF}= 135 \mu$ m,  and $\frac{1}{k_F a_F} \approx 2.6$ where $k_F\approx 0.86 (\mu$m$)^{-1}$ is the Fermi wave vector and $a_F \approx 0.448 \mu$m is the fermionic scattering length at $B=700G$. Noting that $a=0.6 a_F$ \cite{petrov2004}, these parameters give $\lambda= 6.2$ and $g= 3.08 \times 10^4$ in our dimensionless units. 

We find that with these parameters the soliton created after phase imprinting quickly decays into a vortex ring. The period depends sensitively on the minimum ring radius (as in Fig.~\ref{pervrmin}), which in turn depends on the phase imprinting protocol. We can reproduce (within the reported error bars) the experimentally measured period of $T= 4.4 \pm 0.5 T_z$ by relaxing to a vortex ring state using the ansatz in Eq.~\ref{ansatz} with $R_{min}=1.2 a_z$. It is plausible that the particular phase imprinting procedure used in the experiment yields a vortex ring with a similar minimum ring radius. 

\subsection{Ring Dynamics with Axial Asymmetry} \label{secasym}

\begin{figure*}
\centering

\includegraphics[width=1.0\textwidth]{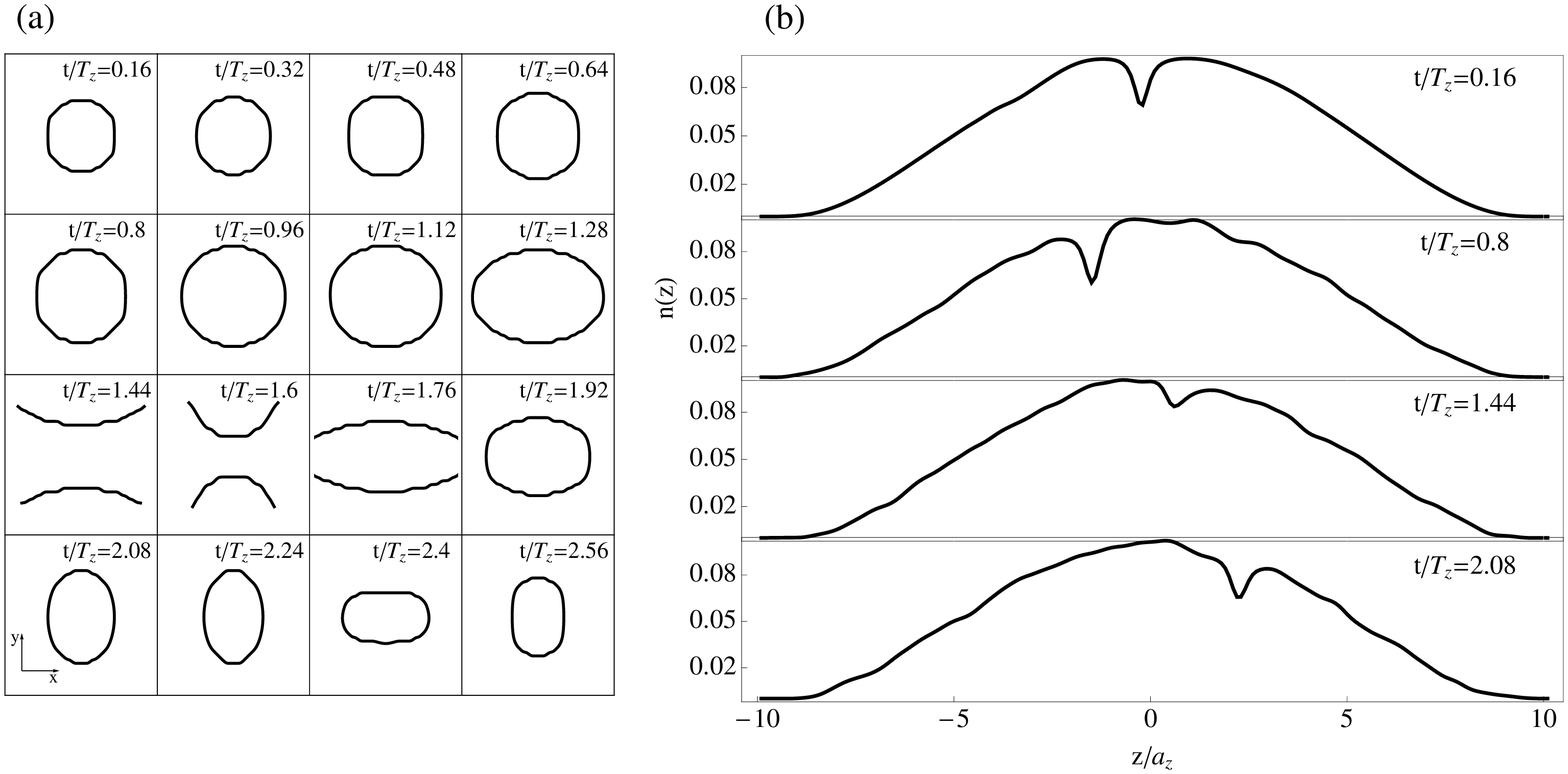}
\caption{Vortex ring dynamics in the presence of a small perturbation to axial symmetry ($V_t(x,y,z)=  \frac{1}{2} [\lambda^2 (0.99x^2+ y^2)+z^2]$). Here $\lambda=4$, $g=4000$, and the system is initialized with a vortex ring of radius $R=1.0 a_z$ located at $z=0$. (a) The locations of non-zero vorticity at different times projected into the $x$-$y$ plane. (b) The density of the condensate at different times integrated over the $x$ and $y$ directions, $n(z)=\int{dxdy \rho(x,y,z)}$.} 
\label{xyplane}
\end{figure*}

To give a more complete picture, we relaxed our assumption of axial asymmetry and performed fully three dimensional simulations with a trap potential given by $V_t(x,y,z)= \frac{1}{2} [\lambda^2 (0.99x^2+ y^2)+z^2]$. We again find the vortex structure moves periodically in the $z$-direction (with roughly the same period), but we find additional internal dynamics, some of which is related to previous studies\cite{horng2006, piazza2011}.

This evolution is illustrated by Fig.~\ref{xyplane}a which shows the locations of non-zero vorticity at different times projected into the $x$-$y$ plane (here $\lambda=4$, $g=4000$). Figure~\ref{xyplane}b shows the density of the condensate integrated over the $x$ and $y$ directions. The system at time $t=0$ contains a vortex ring of radius $R=1.0 a_z$ located at $z=0$. After one half of an oscillation (at $t/T_z\approx 1.4$) the vortex ring breaks apart into two lines of opposite vorticity extending along the $x$ axis, which continue to move together along the $z$ axis. After reaching the edge of the condensate, the vortex lines recombine into a ring which then moves in the opposite direction along the $z$-axis. Similar behavior is seen over the range of parameters explored in Sec.~\ref{secosc}. From the axial density profiles in Fig.~\ref{xyplane}b, however, none of this internal dynamics is apparent. In fact Fig.~\ref{xyplane}b looks like an oscillating grey soliton.

\section{Conclusion} \label{secconc}

Using numerical simulations, we have found that dark solitons created through phase imprinting in three dimensional BECs are unstable to becoming vortex rings, and that these vortex rings oscillate with much larger periods than solitons. We numerically computed the period of these vortex ring oscillations as a function of interaction strength, trap aspect ratio, and minimum vortex ring radius. We found that our results are qualitatively consistent with Jackson et. al.'s \cite{jackson1999} semiclassical model of vortex rings for axially symmetric potential traps. Slight perturbations to axial symmetry produce negligible changes to the oscillation period of the ring, but cause the ring to break apart and recombine during oscillations. Finally, we simulate the BEC regime of a recent experiment claiming to have observed oscillations of ``heavy" dark solitons in cold Fermi gases \cite{yefsah2013}. The oscillation periods of vortex rings in our simulations are quantitatively consistent with the periods of the supposed solitons, and we therefore conclude that these solitons are likely to be vortex rings, or more complicated objects as shown in Fig.~\ref{xyplane}.

A key distinction between vortex rings and solitons, besides their dynamics, is their density profile: a vortex ring appears as two density dips in a two-dimensional profile, while a soliton appears as a solid line of density depletion extending across the condensate. In fermionic superfluids away from the BEC regime, the density depletion associated with vortices and solitons is small, as the cores are filled by normal fluid. For superfluids initially away from the BEC regime, Yefsah et. al. \cite{yefsah2013} were forced to ramp the magnetic field to the deep BEC regime in order to clearly observe \textit{any} density depletion in their gas after releasing it from the trap. We recommend further experiments in the BEC regime where such ramps are unnecessary. We note that previous experiments with BECs have successfully distinguished vortex rings from dark solitons using expansion imaging \cite{anderson2001}, and \textit{in situ} imaging \cite{shomroni2009}. We also note that several of the images in Ref.~\cite{yefsah2013} are suggestive of vortex rings or tangles. This is particularly true of the images in the supplementary information section. 

A less direct distinguishing feature of a vortex ring's dynamics is the asymmetry of its motion. For example if the ring is smaller when moving left to right, it will move faster during that interval than on the return. In Fourier analysis of the existing experimental data, this asymmetry would show up as odd harmonics. We calculated the first odd harmonic for the experimental parameters, and unfortunately found it too small to readily measure. Devising techniques to generate vortex rings with a smaller minimum radii would improve this situation.

Finally, we should mention one shortcoming of our modeling. We find that for our axially symmetric simulations the period of the vortex ring is sensitive to the imprinting protocol,while the experiment finds very reproducible periods. Perhaps the more complicated structures in Sec.~\ref{secasym} yield more reproducibility. The computational cost of the full 3D simulations have prevented us from studying this in detail. 

\section{Acknowledgements}

We thank Martin Zwierlein and Waseem Bakr for discussions and correspondence. We thank Francesco Piazza for making comments leading to our study of the axially anisotropic case. We thank Alexander Fetter for small corrections regarding the two dimensional analog system. This material is based upon work supported by the National Science Foundation Graduate Research Fellowship under Grant No. DGE-1144153 as well as work supported by the National Science Foundation under Grant No. PHY-1068165.

\bibliography{vring}

\end{document}